\documentclass[aps,prfluids,reprint,longbibliography,superscriptaddress,nofootinbib,floatfix]{revtex4-2}

\usepackage{amsmath,amssymb,bm}
\usepackage{graphicx}
\usepackage{booktabs}
\usepackage{microtype}
\usepackage{xcolor}
\usepackage{hyperref}

\newcommand{\lparal}{\ell_{\parallel}}
\newcommand{\lperp}{\ell_{\perp}}
\newcommand{\Dmed}{D_{50}}

\begin{document}

\title{Dynamic Renewal for Hydrodynamic Coupling in Ciliary Mucus Transport}

\author{Jeroen Bechtold}
\affiliation{Department of Information and Computing Sciences, Utrecht University,\\ Princetonplein 5, 3584 CC Utrecht, The Netherlands}

\author{Valentijn van Zwieten}
\affiliation{Department of Information and Computing Sciences, Utrecht University,\\ Princetonplein 5, 3584 CC Utrecht, The Netherlands}

\author{Iris Verkuil}
\affiliation{Department of Information and Computing Sciences, Utrecht University,\\ Princetonplein 5, 3584 CC Utrecht, The Netherlands}

\author{Meike F. Bos}
\affiliation{Institute for Theoretical Physics, Utrecht University,\\ Princetonplein 5, 3584 CC Utrecht, The Netherlands}

\author{Georg Rempfer}
\affiliation{Institute for Computational Physics, Universit{\"a}t Stuttgart,\\ Allmandring 3, 70569 Stuttgart, Germany}

\author{Mae Nesenberend}
\affiliation{Institute for Theoretical Physics, Utrecht University,\\ Princetonplein 5, 3584 CC Utrecht, The Netherlands}

\author{Pieter Michels}
\affiliation{Department of Information and Computing Sciences, Utrecht University,\\ Princetonplein 5, 3584 CC Utrecht, The Netherlands}

\author{Florian K.R. Gaeremynck}
\affiliation{Department of Information and Computing Sciences, Utrecht University,\\ Princetonplein 5, 3584 CC Utrecht, The Netherlands}

\author{Bart Stam}
\affiliation{Department of Information and Computing Sciences, Utrecht University,\\ Princetonplein 5, 3584 CC Utrecht, The Netherlands}

\author{Debabrata Panja}
\affiliation{Department of Information and Computing Sciences, Utrecht University,\\ Princetonplein 5, 3584 CC Utrecht, The Netherlands}

\author{Joost de Graaf}
\email{j.degraaf@uu.nl}
\affiliation{Institute for Theoretical Physics, Utrecht University,\\ Princetonplein 5, 3584 CC Utrecht, The Netherlands}

\date{\today}

\begin{abstract}
\noindent Mucociliary clearance in large mammalian airways involves millimetric MUC5B-rich mucus strands that emerge from submucosal glands approximately aligned with the direction of transport. These strands are observed downstream in cross-flow networks. Recently, Bos~\textit{et al.}~[PNAS Nexus~\textbf{2},~pgad388~(2023)]\nocite{Bos2023} modeled single strand reorientation using a free-draining coupling for the flow induced by cilia. This leaves unresolved whether the strand's own disturbance of the surrounding liquid assists or frustrates the turning. Here, we develop the methodology to address this question. An immersed-boundary coupling is combined with a Fourier-space Stokes solver and a bespoke dynamic-renewal algorithm that exposes the strand to new ciliary domain without enlarging the simulation volume. This allows us to overcome one of the multiscale challenges posed by mucociliary clearance. We find that hydrodynamic interactions result in only a limited decrease in the rate of reorientation, which supports the use of the free-draining approximation. Future work can build on the methodological foundation introduced here to study collective strand dynamics.
\end{abstract}

\maketitle

\section{\label{sec:intro}Introduction}

\noindent Mucociliary transport is a primary mechanical defense of the conducting airways, wherein cilia drive airway-surface liquid and mucus toward the oral cavity, carrying inhaled material out of the lung~\cite{KnowlesBoucher2002, BustamanteMarin2017, hill2022physiology}. In the larger airways of pigs and humans, submucosal glands secrete MUC5B-rich material as long, thick strands rather than as a spatially uniform gel~\cite{WiddicombeWine2015, Ostedgaard2017, Ermund2017, ErmundTransport2018}. These millimetric strands connect into mobile networks and can initiate transport of particles too large to be moved reliably by ciliary contact alone~\cite{Fischer2019}.

Ermund~\textit{et al.}~\cite{Ermund2017} observed that MUC5B strands that emerge from gland ducts are oriented approximately along the direction of mucociliary transport. However, these strands were later found spanning the flow direction and participating in cross-stream networks. They proposed an attachment-based mechanism, in which association with MUC5AC plays a key role. This MUC5AC is produced by and attached to goblet cells, which serve as `sticking points' for the MUC5B strand. Bos~\textit{et al.}~\cite{Bos2023} subsequently demonstrated computationally using a bead-spring model for the MUC5B strand that transient interactions with sparse sticky sites on the surface --- representing MUC5AC strands attached to goblet-cells --- can produce the experimentally observed reorientation. The cilia-induced flow was approximated as uniform and the strand coupled to this using a free-draining approximation. The computational model of Ref.~\cite{Bos2023} also reproduced qualitatively the experimental velocity fluctuations in the MUC5B strands and their slowing down with respect to the flow.

Ramirez-San Juan~\textit{et al.}~\cite{RamirezSanJuan2020} demonstrated that cilia-driven transport is itself heterogeneous across the epithelial surface. Rather than impairing transport, this spatial heterogeneity enhances particle clearance by reducing particle trapping in local recirculating flows and promoting transport across the tissue. Building on this work, Bos~\cite{BosThesis2024} showed that this heterogeneity can also lead to perpendicular configurations of the MUC5B strands, again employing the free-draining approximation within a precomputed flow profile. However, this reorientation occurs on comparatively long travel scales, which supports the importance of MUC5AC attachment in explaining the experimental observations. Both descriptions~\cite{Bos2023, BosThesis2024}, however, leave unresolved whether the strand's local modification of the flow is sufficiently strong to change the reorientation time or statistics.

In creeping flow, the orientation dynamics of elongated particles and filaments are known to depend on geometry, flexibility, confinement, and hydrodynamic interactions~\cite{Jeffery1922, TornbergShelley2004, duRoure2019, Cappello2019}. It is therefore important to resolve hydrodynamic coupling explicitly, because it may significantly impact the dynamics. However, addressing this issue is nontrivial, as this requires overcoming the differences in scale between: (i)~the micron-sized ciliated cells that drive flow; (ii)~the way flow responds to the presence of a millimetric MUC5B strand and \textit{vice versa}; and (iii)~the (multiple) centimeters of upper airways that the strand traverses in reorienting~\cite{Ostedgaard2017, Gilpin2020}.

Bos~\cite{BosThesis2024} relied on lattice-Boltzmann simulations to propagate the coupled dynamics of the strand against a heterogeneous background to establish whether there is a flow-coupling component to reorientation. Without high-performance computing, these could not resolve the full multiscale problem. Even for the free-draining simulation with a precomputed flow profile, the flow profile was restricted to a relatively small simulation box compared to the length of travel required to observe reorientation. To work around this limitation, the strand's direction of travel was taken irrational within the periodic domain, which allowed the strand to cross different parts of the system, whilst limiting the number of times it encountered a similar ciliary-flow-based forcing landscape. However, for sufficiently long reorientation times correlations build up, which interfere with obtaining proper statistics. Employing finite-size corrections proved computationally unfeasible. Therefore, Bos proposed \textit{dynamic renewal} as a means to overcome the multiscale barrier~\cite{BosThesis2024} building on earlier scalable moving window approaches~\cite{Herschlag2019, Puleri2022}. This involves resolving the full dynamics in a suitably chosen environment around the strand, which effectively co-moves with the strand's center of mass (CoM).

Here, we implement this idea and use it to determine whether hydrodynamic coupling measurably changes the distance over which an isolated strand turns when transported by a heterogeneous ciliary field. Our numerical strategy combines standard immersed-boundary force spreading~\cite{peskin1972flow, Peskin2002, ChenPeskin2024}, a spectral solution of periodic Stokes flow using our fast-Fourier-transform-based framework~\cite{rempfer2025projection}, a means to impose boundary constraints, and a moving simulation domain through which the strand is exposed to a pre-generated ciliary texture as it is driven by flow. This design allowed us to directly match two-way-coupled and free-draining simulations and ensure that they encounter the same ciliary-flow-based forcing environment. Comparing various runs, we found that the reorientation statistics agree semi-quantitatively.

\begin{figure}[t]
\centering
\includegraphics[width=0.95\columnwidth]{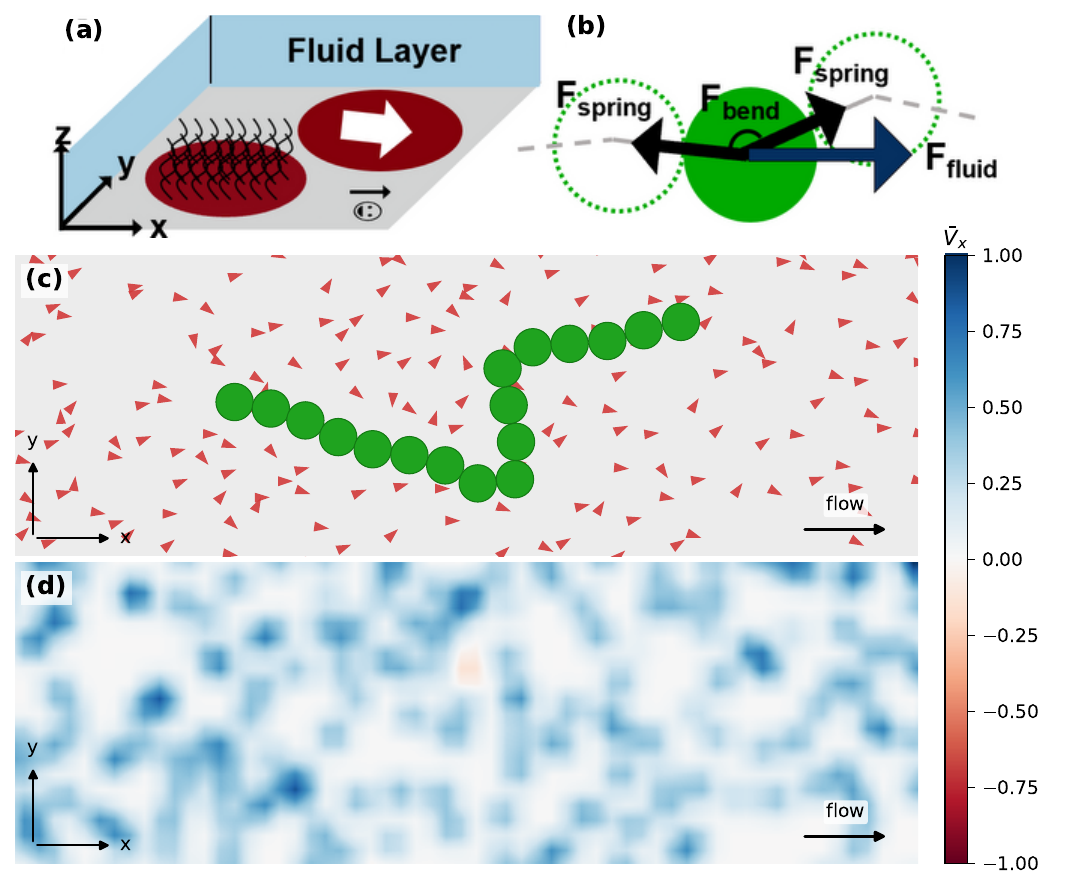}
\caption{\label{fig:system}\textbf{The geometry of MUC5B bundle transport in mucociliary clearance.} (a)~Schematic of our hydrodynamic scenario. The bottom (grey) surface is a no-slip boundary, except for constant-velocity patches (red) representing ciliated cells (sketch) that generate flow along the white arrow. The smiley with black arrow indicates the direction of the oral cavity throughout. (b)~Schematic of the forces acting on a bead in our model mucus strand: linear ($F_{\mathrm{spring}}$) and angular ($F_{\mathrm{bend}}$) springs define its shape and beads are dragged along by the flow ($F_{\mathrm{fluid}}$). (c)~Part of the ciliated surface showing the mucus strand and disordered ciliated-cell patches (red arrows; order parameter $M = 0.81$) to scale. (d)~Magnitude of the normalized flow speed $\bar{V}_{x}$ in the $x$-direction (color bar) close to the ciliated surface. Notice the high-velocity spot close to the kink in the strand shown in (c).}
\end{figure}

\section{\label{sec:mod}Strand, Thin Film, and Liquid}

\noindent Figure~\ref{fig:system}a summarizes our thin fluid-layer geometry and Fig.~\ref{fig:system}b the makeup of our strand. We build on the coarse-grained strand model introduced for this system in Refs.~\cite{Bos2023, BosThesis2024}. That is, we represent a single bundled mucus strand by a bead-spring chain with bead positions $\bm{r}_{i}$ with index $i$ and a total number of $N = 17$ beads. Neighboring beads are connected by finitely extensible nonlinear elastic (FENE) springs~\cite{Warner1972, HerrchenOttinger1997} with potential
\begin{equation}
\label{eq:fene} V_{\mathrm{FENE}} = -\frac{k_{s} \Delta r_{\max}^{2}}{2} \sum_{i=1}^{N-1} \log \! \left[ 1 - \left( \frac{r_{i,i+1} - r_{0}}{\Delta r_{\max}} \right)^{2} \right],
\end{equation}
where $r_{ij} = \vert \bm{r}_{i} - \bm{r}_j \vert$, $r_{0}$ is the equilibrium bond length, $\Delta r_{\max}$ is the maximum deviation from $r_{0}$, and $k_{s}$ is the spring constant. The FENE potential prevents unphysical extension of the coarse-grained strand, while reducing to a harmonic spring for $\vert r - r_{0} \vert \ll \Delta r_{\max}$. The strand's local resistance to bending is represented by
\begin{equation}
\label{eq:bend} V_{\mathrm{bend}} = k_{b} \sum_{i=2}^{N-1} \left(1-\cos\theta_{i}\right),
\end{equation}
with $k_{b}$ the bending coefficient and
\begin{equation}
\label{eq:costh} \cos \theta_{i} = \frac{ ( \bm{r}_{i} - \bm{r}_{i-1} ) \cdot (\bm{r}_{i+1} - \bm{r}_{i}) }{ r_{i-1,i} \, r_{i,i+1} },
\end{equation}
so that a locally straight strand has $\theta_{i} = 0$. Such a bending penalty is required because the secreted bundles remain elongated on transport length scales~\cite{Ermund2017, ErmundTransport2018}. In reduced (grid spacing) units, we take $r_{0} = 2.5$, $\Delta r_{\max} = 2.0$, $k_{s} = 50$, and $k_{b} = 3$, with the latter corresponding to the semi-flexible strand calculations of Ref.~\cite{BosThesis2024}. Note that the persistence length and bending modulus of an entire MUC5B bundle have not been measured experimentally; consequently $k_{s}$ and $k_{b}$ remain phenomenological model parameters. No bead-bead excluded-volume is included in the production runs, since we did not observe overlaps.

The model MUC5B strand is restricted to a thin liquid film with height $H = 3$. This film is bounded by the ciliated surface on the bottom and the liquid-air interface on the top. Every bead experiences a harmonic restoring force toward the mid plane derived from the potential
\begin{equation}
\label{eq:plane} V_{\mathrm{plane}} = \frac{1}{2} k_{w} \left( z_{i} - \frac{z_{\mathrm{top}} + z_{\mathrm{bot}}}{2} \right)^{2},
\end{equation}
where only the beads's $z$-coordinate ($z_{i}$) matters and $z_{\mathrm{top}}$ and $z_{\mathrm{bot}}$ are the position of the top and bottom surface of the thin liquid film, respectively. We chose a stiffness $k_{w} = 50$ throughout, as a practical modeling choice that accounts for unquantified effects of surface tension~\cite{Vanaki2020}. That is, our choice of spring constant is not motivated by biological observation.

Taking the explicit fluid-layer height $H\simeq 75~\mu\mathrm{m}$ and imposed velocity scale $U = 1.3 \simeq 32.5~\mu\mathrm{m}\,\mathrm{s}^{-1}$, as we will justify in Section~\ref{sec:map} and, conservatively, a water-like viscosity $\eta = 10^{-3}~\mathrm{Pa\,s}$ and density $\rho = 10^{3}~\mathrm{kg}\,\mathrm{m}^{-3}$ gives an estimated Reynolds number of
\begin{equation}
\label{eq:Reestimate}\mathrm{Re} = \frac{\rho \, U H}{\eta} \lesssim 2.5 \times 10^{-3}.
\end{equation}
Any increase of the effective airway-fluid viscosity only reduces this value. That is, the fluid inertia may be neglected on the transport scales considered here. We therefore model the suspending liquid as incompressible and Newtonian, using the Stokes equations
\begin{align}
\label{eq:mom} -\bm{\nabla}p + \eta \nabla^{2} \bm{u} + \bm{f} &= \bm{0} ; \\
\label{eq:inc} \bm{\nabla} \cdot \bm{u} &= 0,
\end{align}
where $\bm{u}$ is the fluid velocity, $p$ the pressure, and $\bm{f}$ the external-force density acting on the fluid.

Note that airway surface liquid contains polymeric mucins~\cite{FahyDickey2010} and is not generically Newtonian~\cite{lai2009micro}. However, we choose a simple-fluid constitutive law here, such that we can isolate the effect of two-way hydrodynamic coupling, rather than introduce unknown rheological parameters. The same reduction has been used successfully for cilia-driven airway transport, and the principal effects of spatial heterogeneity were found to persist upon replacing the Newtonian liquid by shear-thinning and shear-thickening generalized-Newtonian fluids~\cite{Vanaki2020, RamirezSanJuan2020}.

\section{\label{sec:flow}Strand Dynamics and Stokes Flow}

\noindent Considering the model MUC5B strand, we note that thermal diffusion is negligible. Consider a particle of diameter $d_{\mathrm{eff}} = 2.48 \simeq 62~\mu\mathrm{m}$, see Section~\ref{sec:map}, within a fluid with the same water-like viscosity as above. At physiological temperature, the particle's Stokes-Einstein diffusivity is given by
\begin{equation}
D_{\mathrm{th}} = \frac{ k_{\mathrm B} T }{ 3 \pi \eta d_{\mathrm{eff}} } \simeq 7 \times 10^{-15}~\mathrm{m^{2}\,s^{-1}} .
\end{equation}
This leads to a conservative bead-scale estimate for the P{\'e}clet number of
\begin{equation}
\label{eq:Peestimate}\mathrm{Pe} = \frac{U d_{\mathrm{eff}}}{D_{\mathrm{th}}} \sim 3 \times 10^{5},   
\end{equation}
and an even larger value when taking the full strand length. Advection therefore overwhelms Brownian translation by many orders of magnitude and $k_{\mathrm B}T = 0$ holds effectively.

The strand dynamics are therefore advanced as follows. For each bead $i$, the elastic, bending, and wall-confinement forces are collected in $\bm{F}_{i}^{\mathrm{int}}$, while the surrounding fluid exerts a Stokes drag proportional to the difference between the local fluid velocity and the bead velocity. Neglecting bead inertia gives the instantaneous force balance
\begin{equation}
\label{eq:beaddynamics} \bm{F}_{i}^{\mathrm{int}} + \zeta \left[\bm{u}(\bm{r}_{i})-\dot{\bm{r}}_{i} \right] = \bm{0} .
\end{equation}
Numerically, Eq.~\eqref{eq:beaddynamics} is integrated with the velocity-Verlet scheme~\cite{Swope1982}, reinstating the inertia term and introducing a bead mass $m = 1$. We set the bead-fluid friction coefficient to $\zeta = 10$. The grid spacing and strength of the coupling give rise to an effective particle radius and accurate fluid-dynamics, as we will return to in Section~\ref{sec:map}. As a rule of thumb, the higher $\zeta$, the stronger the hydrodynamic coupling and the more faithfully hard-sphere interactions are reproduced. However, higher values of $\zeta$ require smaller time steps to ensure numerical stability, as the system becomes increasingly overdamped. Our present choice compromises between these needs.

The velocity $\bm{u}(\bm{r}_{i})$ follows from solving the Stokes problem on a fully periodic Cartesian grid using the discrete projection-based FFT solver of Rempfer~\textit{et al.}~\cite{rempfer2025projection}. In this approach, we enforce discrete incompressibility mode by mode. The zero mode is treated differently, as a periodic Stokes problem has no bounded solution for a nonzero net force. Our system has a net flow imparted by the boundary conditions, but no net force applied on the model MUC5B bundle or the fluid in general. Any spurious net force present on the fluid, due to numerical error, is removed by our projection during the solve step.

In the case of flow-based coupling, we use the immersed-boundary construction of Peskin~\cite{Peskin2002}. Let $\bm{F}_{i}$ be the total force transmitted by bead $i$ to the Newtonian fluid. Then force spreading and velocity interpolation use the same regularized delta function
\begin{align}
\label{eq:ibspread} \bm{f}_{\mathrm{s}}(\bm{x}_{\bm{k}}) &= \sum_{i=1}^{N} \bm{F}_{i} \, \delta_h(\bm{x}_{\bm{k}} - \bm{r}_{i}) ; \\
\label{eq:ibinterp} \bm{u}( \bm{r}_{i} ) &= \sum_{\bm{k}} \bm{u}( \bm{x}_{\bm{k}} )\, \delta_h( \bm{x}_{\bm{k}} - \bm{r}_{i} )\, \Delta x^3 .
\end{align}
Here, $\delta_h(\bm{r}) = \Delta x^{-3} \prod_{\alpha=x,y,z} \phi(r_\alpha/\Delta x)$ is the standard four-point Peskin kernel, with support over four grid nodes in each coordinate direction. Using the same kernel in Eqs.~(\ref{eq:ibspread}) and (\ref{eq:ibinterp}) makes spreading and interpolation discrete adjoints and avoids introducing a separate, inconsistent particle-fluid interpolation rule. Immersed-boundary methods were developed precisely for this type of Lagrangian-Eulerian fluid-structure coupling~\cite{Peskin2002}; more recent Fourier-spectral formulations demonstrate their compatibility with divergence-free Stokes solvers~\cite{ChenPeskin2024}.

\begin{figure}[t]
\centering
\includegraphics[width=0.95\columnwidth]{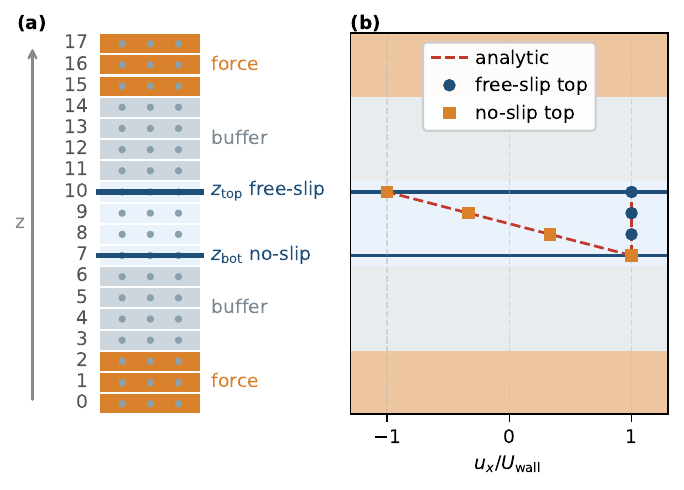}
\caption{\label{fig:bound}\textbf{Boundary geometry and flow validation.} (a)~Vertical cross section of the simulation box showing the FFT grid (dark-gray dots), the rows used for the virtual forces (orange), the buffer zone (gray), and fluid region (light blue). The two blue horizontal lines indicate the position of the bottom and top boundary, where we impose hydrodynamic boundary conditions. (b)~Fluid-velocity profiles for two sets of boundary conditions. (i)~A uniform velocity applied to the no-slip bottom and top surface --- the top surface moves in the opposite direction from the bottom one --- leading to a Couette flow profile (orange squares). (ii)~A no-slip, uniform velocity condition is applied on the bottom surface, paired with a free-slip condition on the top surface, leading to a uniform velocity profile (blue circles). Both profiles are compared to the analytic theory (dashed, red lines).}
\end{figure}

Physical boundaries are embedded in the periodic solver to account for the ciliated surface over which the strand moves and the liquid-air interface. These are accounted for by layers of virtual forces outside the liquid region, in the spirit of fixed-grid forcing methods for no-slip boundaries~\cite{Goldstein1993, MittalIaccarino2005, Uhlmann2005}, as we will return to in Section~\ref{sec:bound}. The typical production grid is $N_{x} \times N_{y} \times N_{z} = 128 \times 96 \times 18$, which we tested to limit periodicity artifacts. We use three virtual-force rows separated from top and bottom boundary by four buffer rows\footnote{The Peskin kernel sits at a grid node, hence it spreads forces onto one row directly below and above the node of application. We therefore expect to require at least one buffer row to contain the tail of the Peskin spread of the virtual forces. Otherwise we would apply virtual forces inside the fluid domain and change the physics. The remaining buffer rows were found to be necessary to provide the pseudoinverse enough freedom to satisfy the wall boundary condition and the zero-net-force constraint for the entire fluid. Four buffer rows, plus three virtual-force rows proved to be the smallest stable configuration that passed both the conditioning test for the cilia boundary condition, as well as the 0.1\% deviation from linearity criterion imposed on Couette profile.}, thus placing ciliated surface and liquid-air interface at $z_{\mathrm{bot}} = 7$ and $z_{\mathrm{top}} = 10$, respectively, which leads to the value of $H = 3$ above.

The lower boundary represents the time-averaged action of the ciliated epithelium. We do not resolve individual ciliary strokes. Instead, the bottom wall carries a coarse-grained heterogeneous tangential velocity field constructed from effective ciliary sites, following the approach proposed by Ram{\'i}rez-San Juan~\emph{et al.}~\cite{RamirezSanJuan2020}. As we will discuss in Section~\ref{sec:DR}, this follows from a pre-generated domain. On this domain, which is not periodic and larger than the simulation box, candidate sites are placed randomly on a square lattice of spacing $s = 2.20$, and each site is independently active with probability $\phi_{c} = 0.65$. This gives an average model multiciliated-cell surface coverage that lies in the range found for porcine airways~\cite{wallace1994transdifferentiation}. Figure~\ref{fig:system}c shows an example selection of site positions.

For an active site $j$, the in-plane direction $\theta_j$ is drawn from the von-Mises distribution~\cite{MardiaJupp1999, RamirezSanJuan2020}
\begin{equation}
\label{eq:vonmises} p(\theta) = \frac{ \exp[ \kappa \cos(\theta-\theta_{0})] }{ 2 \pi I_{0}(\kappa) } ,
\end{equation}
with $\theta_{0} = 0$ and $\kappa = 3$, where $I_{0}(\kappa)$ is the modified Bessel function of the first kind of order 0. Here, we have chosen the $x$ direction as the mean transport direction, also see Fig.~\ref{fig:system}c. The corresponding orientational order is
\begin{equation}
\label{eq:M} M = \left\vert \left\langle e^{i\theta} \right\rangle \right\vert = \frac{ I_1(\kappa) }{ I_{0}(\kappa) } \simeq 0.81 ,
\end{equation}
with $I_{1}$ the order 1 version of the above modified Bessel function. The porcine value of $M$ is not known; $M=0.81$ follows the previous model~\cite{BosThesis2024} and matches the tissue-scale value $M = 0.80 \pm 0.13$ measured in mouse trachea~\cite{RamirezSanJuan2020}. Each active site supplies an effective velocity $U_{c}( \cos \theta_{j}, \sin \theta_{j})$ with $U_{c} = 11.9$ in lattice velocity units\footnote{This follows from inverting Eq.~\eqref{eq:umean} with $\phi_{c} = 0.65$, $s = 2.20$, and $M = 0.81$, each unit of $U_{c}$ adds $0.109$ to the mean wall speed, so $U_{c} = 1.30 / 0.109$.}, which is spread onto the wall using the two-dimensional (2D) version of the four-point Peskin kernel. The boundary field reads
\begin{equation}
\label{eq:ciliawall} \bm{u}_{\mathrm{wall}}(x,y) = \sum_{j} U_{c} \begin{pmatrix} \cos\theta_{j} \\ \sin\theta_{j} \end{pmatrix} \delta_h^{(2)} \left( \begin{pmatrix} x \\ y \end{pmatrix} -  \bm{p}_{j} \right), 
\end{equation}
with $p_{j}$ the 2D center point of the active sites and $\delta_{h}^{(2)}$ the 2D variant of the four-point Peskin kernel. The expected streamwise mean speed is
\begin{equation}
\label{eq:umean} \left\langle u_{\mathrm{wall},x} \right\rangle \simeq \frac{\phi_c}{s^{2}} U_{c} \frac{I_{1}(\kappa)}{I_{0}(\kappa)} \simeq 1.30 ,
\end{equation}
which we confirmed by explicit calculation, see Fig.~\ref{fig:system}d for the flow field associated with an example ciliated surface.

The upper boundary represents the liquid-air interface. We impose the boundary condition
\begin{equation}
\label{eq:topbc} \left. \partial_{z} \bm{u} \right\vert_{z=z_{\mathrm{top}}} = 0,
\end{equation}
so that the tangential components satisfy the zero-shear condition appropriate to a liquid bounded by a much less viscous gas; no penetration $u_{z} = 0$ ensures no fluid-mass transport across the interface. The same idealization is used in cilia-driven airway-film calculations~\cite{RamirezSanJuan2020}.

\section{\label{sec:bound}Boundary-Condition Algorithm}

\noindent The boundary conditions are imposed through virtual forces placed on lattice sites outside the physical channel, see Fig.~\ref{fig:bound}a. Let $\bm{u}^{\mathrm{str}}(x,y)$ denote the velocity field obtained from the the interpolated forces acting from the strand onto the fluid before the wall correction is applied. We collect the boundary conditions in a vector $\bm{\mathcal{B}}$ as
\begin{equation}
\bm{\mathcal{B}}(\bm{u}) = \begin{pmatrix} \bm{u}(x,y,z_{\mathrm{bot}})\\[1mm] \partial_{z} \bm{u}(x,y,z_{\mathrm{top}}) \end{pmatrix} ,
\end{equation}
which represents a vector with 6 components. The prescribed wall values are given by
\begin{equation}
\bm{b} = \begin{pmatrix} u_{\mathrm{wall},x}(x,y) & u_{\mathrm{wall},y}(x,y) & 0 & 0 & 0 & 0 \end{pmatrix}^{\mathsf{T}} ,
\end{equation}
where $\mathsf{T}$ denotes transposition. Thus, the wall residual that must be canceled by the virtual forces is
\begin{equation}
\label{eq:wallresidual} \bm{r}_{\mathrm{w}} = \bm{b} - \bm{\mathcal{B}}\left(\bm{u}^{\mathrm{str}}\right).
\end{equation}
Here, the first three components enforce the prescribed cilia-driven velocity at the lower wall, whereas the latter three enforce the zero-normal-gradient condition at the upper wall. The derivatives are understood as the corresponding discrete derivatives of the spectral solver~\cite{rempfer2025projection}.

Because both the Stokes solver and the immersed-boundary coupling are linear, the response of the fluid at the boundaries to (virtual) forces can be written as a linear operator $\underline{\bm{A}}$. This includes the Peskin spread of forces from the ghost rows to the Eulerian grid and the Stokes solve. Consequently, we can cast the problem of obtaining the appropriate virtual forces as 
\begin{equation}
\label{eq:wallrealspace} \bm{r}_{\mathrm{w}} =  \underline{\bm{A}} \, \bm{f}_{\mathrm{virt}} ,
\end{equation}
with $\bm{f}_{\mathrm{virt}}$ a generalized vector containing all virtual forces.

An element of the matrix $A_{\alpha\beta}$ thus gives the change in boundary constraint $\alpha$ --- a velocity component at the lower wall or a velocity-gradient component at the upper wall --- generated by a unit virtual-force degree of freedom $\beta$. For the production geometry, three virtual-force rows are used on either side of the channel, giving $18$ virtual-force components per lateral position and six physical boundary constraints. The system is therefore rectangular and underdetermined. We solve it in the least-squares sense using the Moore-Penrose pseudoinverse $\underline{\bm{A}}^{+}$~\cite{penrose1955generalized}. Additional zero-mode constraints required by the discrete spectral solver may be appended to this system before the pseudoinverse is constructed, if needed.

Computing the pseudoinverse of the full real-space operator $\underline{\bm{A}}$ proved computationally prohibitive because it involves all $N_{x}N_{y}$ lateral positions. However, both the channel geometry and the Peskin spreading kernel are translationally invariant in $x$ and $y$. Hence $A_{\alpha\beta}$ depends on the lateral coordinates only through the separation between the corresponding wall and force nodes, making $\underline{\bm{A}}$ block circulant in the two lateral directions. A 2D discrete Fourier transform~\cite{Trefethen2000} --- written using check notation~\cite{rempfer2025projection} --- therefore diagonalizes these lateral convolutions and reduces Eq.~\eqref{eq:wallrealspace} to one independent system for every lateral wave-vector,
\begin{equation}
\label{eq:wallfourier} \check{\bm{f}}_{\mathrm{virt}}(k_{x},k_{y}) = \check{\underline{\bm{A}}}^{+}(k_{x},k_{y}) \, \check{\bm{r}}_{\mathrm{w}}(k_{x},k_{y}).
\end{equation}
The matrices $\check{\underline{\bm{A}}}(k_{x},k_{y})$ depend only on the fixed grid, wall geometry, Peskin kernel, and discrete Stokes operator. The (relatively small) reduced Moore-Penrose pseudoinverse can therefore be computed once at initialization. Each subsequent wall correction requires only a 2D FFT of the residual, independent matrix-vector products for the Fourier modes and an inverse FFT of the resulting virtual forces.

Note that this translation invariance applies to the \textit{response operator}, not to the imposed wall velocity.  A unit virtual force produces the same wall response after any common translation of source and observation point. The heterogeneous ciliary velocity instead enters through the residual
$\bm{r}_{\mathrm w}(x,y)$ and therefore through its Fourier amplitudes. Consequently, the block-circulant operator can be diagonalized even for an arbitrarily heterogeneous prescribed wall velocity: each lateral Fourier component of that velocity is simply acted on by the corresponding small matrix $\check{\underline{\bm{A}}}^{+}(k_{x},k_{y})$.

We found that our spectral, Moore-Penrose pseudoinverse solver is capable of accurately reproducing analytic fluid profiles. We tested a Couette flow, imposing a uniform bottom-boundary velocity and oppositely directed uniform top-boundary velocity (both no-slip), as well as the same situation where the top boundary is replaced by a free-slip surface. For both scenarios, we compared the numerically obtained velocity profiles against the ones expected from analytic theory,~\textit{i.e.}, linear in the case of the Couette flow and constant with the free-slip surface, see Fig.~\ref{fig:bound}b. We observed a maximum velocity-field difference of $2.5 \times 10^{-14}$ in reduced unit over the four lattice layers that contribute to the fluid, which indicates that our solver can achieve machine-precision accuracy. 

\section{\label{sec:DR}Dynamic Renewal}

\noindent A spectral solver is forced to use a small periodic box based on memory availability and accuracy requirements. Here, we employ the length $\simeq 3.2~\mathrm{mm}$, which was found to empirically to balance performance with accuracy.  However, reorientation was observed to occur in simulations over centimeters without the addition of surface adhesion~\cite{BosThesis2024}. To overcome the issues of size, we pre-generate, for each random seed, a non-periodic ciliary velocity boundary-condition domain. This pre-generated domain is chosen to be longer than the entire strand trajectory, $10~\mathrm{cm}$ here, and wider than the simulation box. The velocity boundary is computed using the rules provided in Section~\ref{sec:flow}.

\begin{figure}[t]
\centering
\includegraphics[width=0.95\columnwidth]{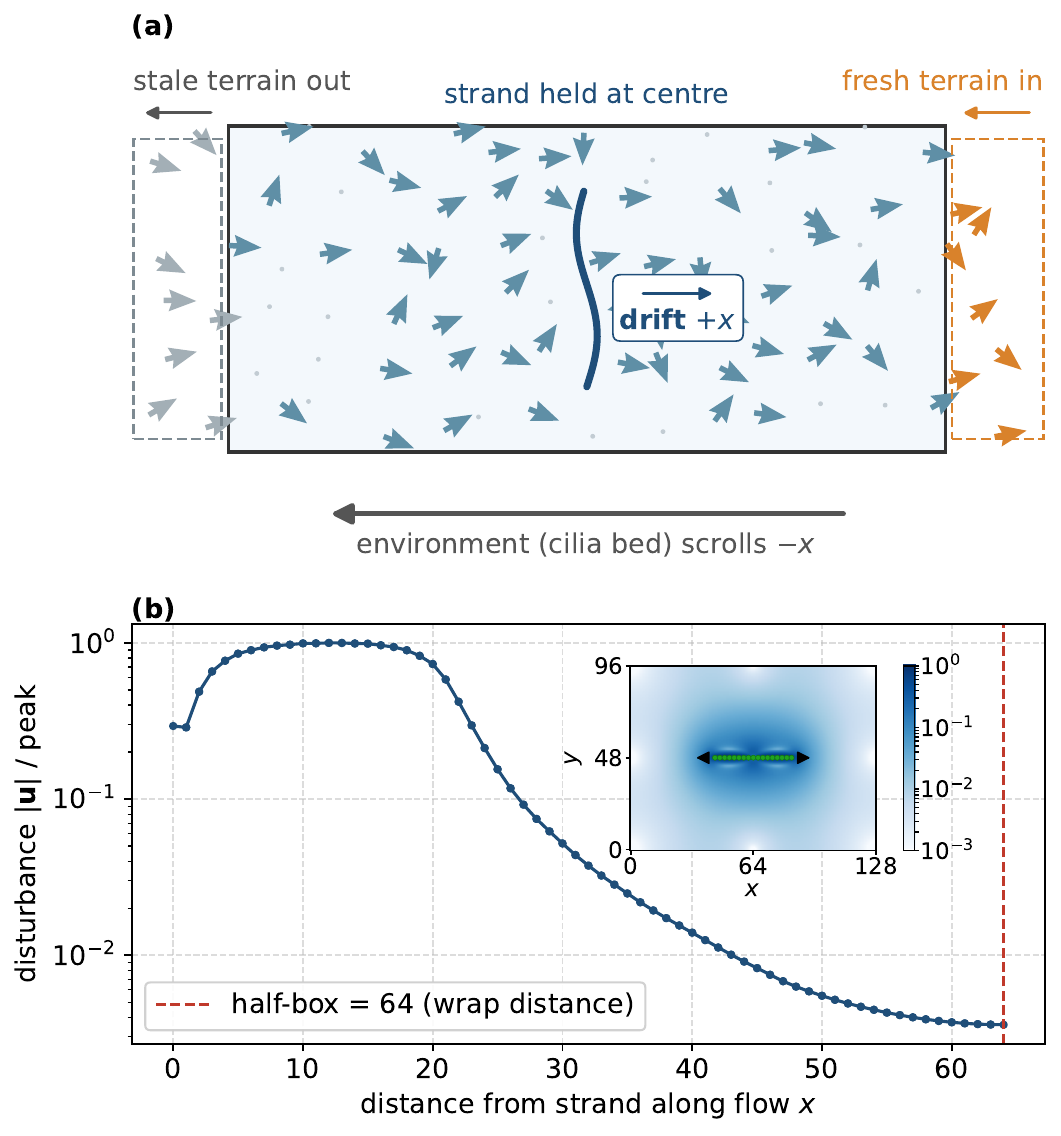}
\caption{\label{fig:method}\textbf{Dynamic renewal approach.} (a)~Illustration of the central simulation box being periodically supplied with unexplored ciliated domain as the strand advances, while passed domain is clipped. (b)~Decay of a strand’s velocity disturbance along the flow direction (see inset). The maximum velocity $\vert \bm{u} \vert$ peaks near the strand and falls by more than two orders of magnitude, reaching 0.36\% of its peak at the
half-box distance $N_{x}/2 = 64$ (dashed vertical line).}
\end{figure}

When the strand's CoM drifts by a prescribed threshold, the bead coordinates are relabeled back to the box center and the wall-velocity domain is advanced by the same accumulated displacement. That is, part of the unexplored epithelial domain enters at the leading edge and a previously encountered, equal-size part of the domain leaves at the trailing edge. In our simulations we found eight lattice sites for the drift ($\simeq 0.2~\mathrm{mm}$) to give a good balance between frequency of renewal and accuracy. Note that no random numbers are drawn during these updates, which reduces the computational cost. Figure~\ref{fig:method}a shows our approach. Importantly, here we show dynamic renewal in the length-wise direction, as our strand did not drift appreciably in the sideways direction. However, the method can be adjusted to allow for this as well.

This construction has three useful consequences. First, it separates the periodicity of the fluid solver from the biological forcing. The fluid remains periodic, but the strand never sees a repeated ciliary patch; the box is sufficiently large to suppress edge effects, as we will return to shortly. Second, paired free-draining and hydrodynamic-coupling simulations can be matched. That is, for a given seed, both experience exactly the same ciliary domain and initial strand configuration. The only difference is whether the strand force is included in the Stokes solve, which leads to a divergent trajectories on the long term. Third, finite-size scaling becomes computationally feasible. The resolved box is relatively small, thus it can be varied to probe the effect of periodicity and the edges.

The dynamic-renewal operation was tested directly on the flow resulting from the combined system of a strand in a non-uniform ciliary landscape. Across window shifts, the velocity jump in the flow field close to the strand was found to be below $4 \times 10^{-15}$. Performing a sweep over a set of shift sizes revealed that the strand's fluid wake, which is not refreshed, had a small impact on the dynamics. The measured streamwise disturbance from a forced strand fell by more than two orders of magnitude and reached $0.36\%$ of its peak value at the half-box distance where the image wraps, see Fig.~\ref{fig:method}b. These checks support that we can use dynamic-renewal updates to capture the model MUC5B strand's propagation over the model ciliated surface.

\section{\label{sec:map}Parameter Mapping}

\begin{figure}[t]
\centering
\includegraphics[width=0.75\columnwidth]{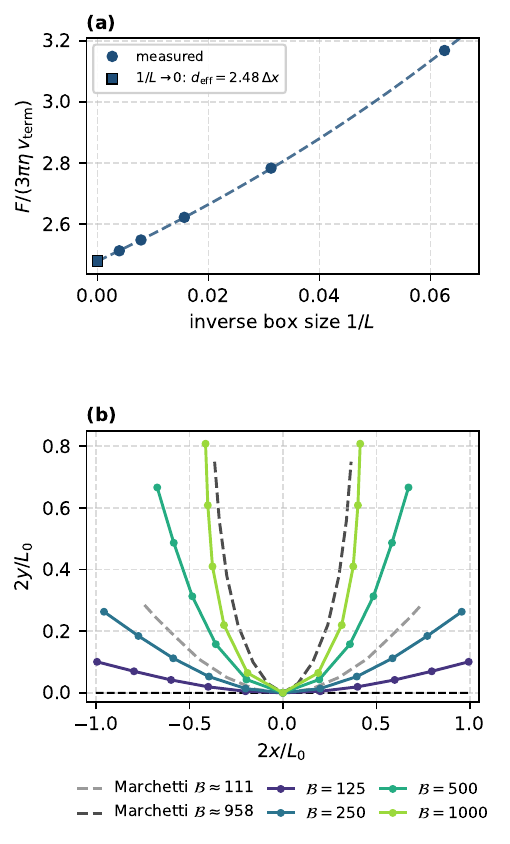}
\caption{\label{fig:bench}\textbf{Parameter mapping and validation.} (a) The ratio $F/(3 \pi \eta v_{\mathrm{term}})$, with $v_{\mathrm{term}}$ the terminal velocity and $F$ the applied force, as a function of the inverse box length $1/L$. The dashed curve fits the expected Hasimoto scaling to the data (dots) obtained for a single bead moving in a periodic 3D box. The extrapolated value matches the effective bead diameter $d_{\mathrm{eff}}$ in lattice units (square). (b) Shape of a single test fiber comprising $N = 11$ beads (and associated size $L_{0} \approx 25$) settling under the force of gravity $f_{g}$ in 3D. Several values of the elasto-gravitational number $\mathcal{B} = f_{g} L_{0}^{2} / (k_{b} d_{\mathrm{eff}})$ (dots) are shown and compared to data from Marchetti~\textit{et al.}~\cite{Marchetti2018} (dashed gray and black curves). The dashed horizontal line at $y = 0$ represents the limit of a rigid rod.}
\end{figure}

\noindent We used a Hasimoto finite-size extrapolation~\cite{Hasimoto1959, Nash2008, fischer2015raspberry} to determine the effective bead diameter of $d_{\mathrm{eff}} = 2.48$ for the four-point Peskin kernel, see Fig.~\ref{fig:bench}a. This choice ensured that the mobility of the bead changed by only $0.14\%$ over a sub-grid translation.

After obtaining $d_{\mathrm{eff}}$ a parameter mapping is possible. Experimental observations on MUC5B bundles form the reference and inform our parameter mapping. Airway bundles can extend to millimetric lengths~\cite{Ostedgaard2017, Ermund2017, ErmundTransport2018}; we therefore used $1~\mathrm{mm}$ as the model strand's contour length~\cite{Bos2023, BosThesis2024}. To account for the measured bead diameter, we chose $r_{0} = 2.5 \simeq d_{\mathrm{eff}}$ so that adjacent hydrodynamic blobs approximately touch rather than forming a chain of well-separated point particles.

Using these choices additionally verified hydrodynamic coupling within a strand by studying a test fiber with $N = 11$ beads. This is sedimented under the influence of a homogeneous gravitational field in a 3D periodic domain. Its steady-state shape attained the expected U-shaped deformation, see Fig.~\ref{fig:bench}b. We also found that the normalized settling speed grows with flexibility~\cite{Marchetti2018}. However, the magnitude of the speed increase is smaller than in the slender-fiber reference, because the effective aspect ratio of the test strand is smaller (about 10 compared to between 51 and 303 in Ref.~\cite{Marchetti2018}), resulting in reduced drag anisotropy. The fact that we were able to capture the U-shaped deformation indicates that our choice of $\zeta$ accounts for (steady-state) hydrodynamic interactions.

Returning to our model MUC5B strand, the relaxed contour length is $L_{0} = (N-1)r_{0} = 40$. Identifying $L_{0}$ with $1~\mathrm{mm}$ fixes our unit length $\simeq 25~\mu\mathrm{m}$. This leads to an effective hydrodynamic diameter of $d_{\mathrm{eff}} \simeq 62~\mu\mathrm{m}$. This value is slightly larger than the $20$ to $30~\mu\mathrm{m}$ typically observed in experiments~\cite{Ermund2017, ermund2018mucus}, or the broader range of approximately $5$ to $50~\mu\mathrm{m}$ reported by~\cite{WiddicombeWine2015, Ostedgaard2017,  ErmundTransport2018}. The point of our present calculation is a proof of principle, thus the larger width than typically observed is not a significant issue. Given that $H = 3 \simeq 75~\mu\mathrm{m}$, the relevant confinement ratio is $H/d_{\mathrm{eff}} \simeq 1.21$,~\textit{i.e.}, the strand occupies an order-unity fraction of the film depth, qualitatively consistent with the strong vertical confinement experienced by airway MUC5B bundles. In biological experiments, strands were observed to be thicker than the confining fluid layer~\cite{ermund2018mucus}, which would be challenging to model in our FFT setup. We further compute that $\left\langle u_{\mathrm{wall},x} \right\rangle \simeq 1.30 \simeq 32.5~\mu\mathrm{m}\,\mathrm{s}^{-1}$, in line with the flow speeds reported in the biological literature~\cite{Liu2013}, though it is a factor of 3 to 4 below the speed reported in Ref.~\cite{hoegger2014assessing} and above that reported in Ref.~\cite{RamirezSanJuan2020}, respectively.

\section{\label{sec:impl}Implementation Details}

\noindent The simulations were implemented in \textbf{\texttt{Python}} using \textbf{\texttt{JAX}}~\cite{JAX2018} and executed on a single NVIDIA GeForce RTX 4070 GPU with $12~\mathrm{GB}$ of device memory, an AMD Ryzen 7 3800X CPU, and $32~\mathrm{GB}$ of system memory. The time-stepping loop was JIT-compiled through XLA and executed in chunks using \textbf{\texttt{jax.lax.scan}}, thereby avoiding host-device synchronization at every time step. We used a reduced time step of $\Delta t = 10^{-3}$, corresponding to approximately $10^{-3}~\mathrm{s}$ in physical units. For the production grid of $128 \times 96 \times 18$, a fully coupled time step required approximately $5.4~\mathrm{ms}$ of computational time in profiling, while the complete production loop, including conveyor-belt updates and data output, averaged approximately $6.2~\mathrm{ms}$ per step over the 50 coupled production runs, also see Table~\ref{tab:performance}. All production calculations were performed in double precision by enabling  \textbf{\texttt{JAX}}'s 64-bit mode. The materials required to reproduce the results presented here are provided in the accompanying open data package~\cite{BechtoldData2026}.

\begin{table}[t]
\centering
\caption{Computational performance on the production grid ($128\times96\times18$). The first three values are obtained from profiling, whereas the complete production loop includes conveyor-belt updates and data output.}
\label{tab:performance}
\begin{tabular}{lr}
\toprule
Measurement & Time (ms/step) \\
\midrule
Free-draining time step       & 1.89 \\
Fully coupled time step       & 5.36 \\
Additional cost of coupling   & 3.47 \\
Complete production loop      & 6.17 \\
\bottomrule
\end{tabular}
\end{table}

Profiling showed that the computational cost is dominated by the spectral fluid solver; the Lagrangian strand operations contribute far less to these. Immersed-boundary spreading of the bead forces required approximately $0.3~\mathrm{ms}$ per fluid evaluation, whereas the wall-corrected Stokes solve was the most expensive component. JIT compilation further reduces the effective computational cost by moving operations that remain constant within a scan chunk outside the time-stepping loop. This is particularly advantageous in the free-draining calculation, where the cilia-driven flow remains fixed within a chunk and the complete fluid solution can be reused. In the fully coupled calculation, the strand forces change at every step, requiring repeated evaluation of the spectral fluid solver and accounting for most of the increase from approximately $1.9~\mathrm{ms}$ to $5.4~\mathrm{ms}$ per step.

\section{\label{sec:data}Data Collection and Analysis}

\noindent We use 50 shared random seeds to determine the impact of hydrodynamic coupling. That is, every seed is run once with full two-way coupling (C) and once in the free-draining (FD) limit. During a run, we keep track of the normalized streamwise and cross-stream extents of a strand's configuration, $\lparal$ and $\lperp$, respectively. Following the earlier mucus-strand work \cite{Bos2023, BosThesis2024}, a strand is classified as reoriented at the first traveled distance $d_{i}^{a}$ for which $\lperp > \lparal$, where $i$ indexes the seed and $a$ the dynamical model. In other words, the strand is reoriented when the effective bounding-box angle exceeds $45^{\circ}$. Each realization is followed until the strand has displaced $L = 10~\mathrm{cm}$, which was chosen based on the calculations of Ref.~\cite{BosThesis2024}.

We describe the ensemble through the cumulative reorientation curve $F_{a}(d)$, which gives the fraction of strands that have reoriented by traveled distance $d$. For visualization we fit our empirical curve with the two-parameter logistic form
\begin{equation}
\label{eq:logisticfit} F_{a}^{\mathrm{fit}}(d) = \frac{1} {1+\exp[-(\beta_{0,a}+\beta_{1,a}d)]} .
\end{equation}
This allows us to extract a median reorientation distance as the $50\%$ crossing of this fitted curve, $\Dmed^{a} =-\beta_{0,a} / \beta_{1,a}$. The logistic fit is used here only to provide a smooth and reproducible estimate of this characteristic distance.

\begin{figure}[t]
\centering
\includegraphics[width=0.95\columnwidth]{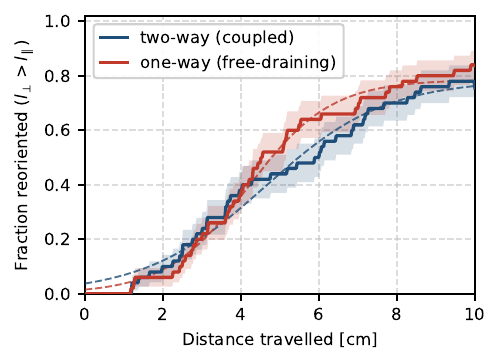}
\caption{\label{fig:result}\textbf{The role of hydrodynamic coupling.} Fraction of the 50 simulated strands that have reoriented perpendicular ($\lperp > \lparal$) to the flow as a function of the traversed distance $d$; here in centimeters. The two-way coupled strand data is in blue, while that obtained using the free-draining approximation is given in red. Solid lines indicate the mean and the bands the standard error of the mean. The dashed curves represent logistic fits to the mean data using Eq.~\eqref{eq:logisticfit}.}
\end{figure}

Figure~\ref{fig:result} contains our comparison; the open data package~\cite{BechtoldData2026} contains videos showing individual trajectories. The fraction of strands that have reoriented by traveled distance $d$ follows nearly the same curve with and without coupling. The fully coupled ensemble has $\Dmed = 4.6~\mathrm{cm}$, while the free-draining ensemble gives $4.2~\mathrm{cm}$. At the common $10\,\mathrm{cm}$ cutoff, 39 of 50 coupled strands and 42 of 50 free-draining strands have reoriented. Thus, the complete distance-to-reorientation curves are statistically consistent with a common event rate (log-rank $p=0.48$)~\cite{Jung1999}. We conclude that including hydrodynamic coupling in this system does not significantly impact the reorientation time; individual comparisons further reveal a limited effect on the overall dynamics.

\section{\label{sec:discuss}Discussion}

\noindent It is useful to distinguish the two reductions used here. Dynamic renewal reduces the \textit{spatial} problem, while the free-draining approximation instead reduces the \textit{physics}. Dynamic renewal makes it practical to test the latter approximation over long travel distances. Our approach shares the locality and scale-separation principle of heterogeneous multiscale methods~\cite{EEngquist2003, Abdulle2012HMM}. However, it is algorithmically closer to moving-window or adaptive-physics-refinement approaches, which retain detailed physics only where it is needed~\cite{Herschlag2019, Puleri2022}.

Dynamic renewal's computational savings stem from discarding fluid that is far from the object. Thus, the condition for this construction to work is that any information left behind by the moving window must no longer significantly influence the local dynamics. This is not automatic in Stokes flow, due to the typically long-ranged nature of hydrodynamic flows\footnote{In 3D, the leading hydrodynamic interaction term to decay $\propto r^{-1}$, while near a single no-slip boundary it is $\propto r^{-3}$ and near a free-slip boundary it remains $\propto r^{-1}$. For fluids confined between two parallel no-slip walls and for a no-slip / free-slip combination we find $\propto r^{-2}$, and for two free-slip surfaces there is logarithmic decay $\propto \log L / r$ with $L$ the extent of the plates. Note that momentum is exponentially screened for two parallel no-slip walls, because these walls absorb momentum, but incompressibility preserves a long-ranged pressure-driven flow that is $\propto r^{-2}$.}. However, in our confined geometry, the strand-induced disturbance has fallen to $0.36\%$ of its peak value before reaching the half-box distance, which we considered acceptable. The same assumption would need to be checked in other systems. Likewise, the present renewal scheme assumes that the surface itself has no long-lived response to the passing object. Deformable, adhesive, or otherwise history-dependent boundaries, all of which can be encountered in the upper airways~\cite{Hoegger2014, Ermund2017, ErmundTransport2018, Loiseau2020}, would require additional states to be carried with the moving window and separately checked for their influence.

Given these caveats, we find that the payoff of using dynamic renewal can be substantial. Extending the FFT domain over the full $10~\mathrm{cm}$ trajectory would require $N_{x} = 4096$ and about $295~\mathrm{ms}$ per coupled step (estimated using the $\mathcal{O}(N \log N)$ scaling), whereas the renewed $N_{x} = 128$ domain requires about $5.4~\mathrm{ms}$. Future work will focus on improving the efficiency of the algorithm further by reducing the number of buffer rows needed for the Peskin interpolation of the virtual forces.

The idea of dynamic renewal could find application in systems where transport is known to depend on spatially resolved surface structure. Leukocyte rolling gives an example of such a system~\cite{Herman2011, Lee2011}. The rolling velocity and attachment vary strongly with local P-selectin density, and micrometer-scale P-selectin patterns can laterally steer cells as they move under flow. In our approach, a measured or prescribed endothelial ligand map could be moved past a locally resolved cell-flow calculation, rather than resolving an entire vessel at cellular resolution. This resembles aspects of the scale separation in adaptive physics refinement, where a high-resolution fluid-structure window is moved with a circulating cell through a much larger vascular domain~\cite{Herschlag2019, Puleri2022}. Active colloids provide a second example. Janus microswimmers have been shown to dock at and follow sub-micrometer surface steps, with their reorientation controlled by phoretic and hydrodynamic interactions with the nearby topography \cite{Simmchen2016}. In the same study, the influence of the step was found to be negligible beyond roughly three swimmer radii. This has the requisite structure required for dynamic renewal.

\section{\label{sec:close}Summary and Outlook}

\noindent In summary, we have introduced a moving-boundary strategy, which we refer to as \textit{dynamic renewal}, for studying multiscale transport processes in airway clearance over distances far exceeding the computational domain. It is a variant of traditional multiscale methods, tailored to studying model MUC5B strand dynamics. We applied this approach to investigate such a strand's reorientation in cilia-driven flow, where we compared fully coupled dynamics, in which the strand modifies the surrounding fluid, with a free-draining approximation that neglects this hydrodynamic coupling.

Flows are computed using an FFT-based Stokes solver with immersed-boundary coupling (for a coupled strand) and virtual boundary conditions. Dynamic renewal prevents the translating strand from repeatedly encountering the same realization of the ciliary surface. This removes a finite-domain artifact that had previously limited long-distance, fully coupled mucus-strand simulations~\cite{BosThesis2024}. For the isolated strand and parameter regime considered here, resolving hydrodynamic coupling produced only a small change in the reorientation-distance distribution. Our result thus provides support for the use of the substantially cheaper free-draining approximation in earlier work.

Looking forward, we can apply dynamic renewal to study multiple interacting strands, mucus-network formation, and it interaction with inhaled debris~\cite{ErmundTransport2018, Fischer2019}.

\begin{acknowledgments}
\noindent J.d.G. acknowledges NWO for funding through Start-Up Grant No. 740.018.013. We would also like to thank Dr.~Anna Ermund and Prof.~Gunnar C.~Hansson for useful discussions in setting up our original modeling~\cite{Bos2023, BosThesis2024}.

Realizing the dynamic-renewal algorithm using an FFT solver involved several student projects. The author contributions are as follows: Conceptualization, J.d.G.; methodology, J.B., M.B., J.d.G. (dynamic renewal), J.B., V.v.Z, I.V., and P.M. (efficient pseudoinverse solving in complex geometries), M.N. (incompressibility condition), F.K.R.G., B.S., D.P., and J.d.G. (understanding and improving boundary conditions), G.R. (original FFT approach), M.B. (original strand model); numerical calculations, J.B.; validation, all; investigation, all; writing—original draft, J.d.G. and J.B.; writing-review and editing, J.B. and D.P.; funding acquisition, J.d.G.; resources, J.d.G.; supervision, J.d.G. and D.P.
\end{acknowledgments}

\section*{Data availability}
\noindent The data that support the findings of this article are openly available in Ref.~\cite{BechtoldData2026}. This contains the codes required to reproduce the numerical experiments and validations, the data for the longer runs, as well as scripts to produce the figures from processed data and the supplemental movies underlying some of the data in Fig.~\ref{fig:result}.

\bibliography{dynamic_renewal}

\end{document}